# TOWARDS CLOUD COMPUTING: A SWOT ANALYSIS ON ITS ADOPTION IN SMES

Kimia Ghaffari[1], Mohammad Soltani Delgosha[1] and Neda Abdolvand[1]

[1]Alzahra University, Tehran, Iran

## ABSTRACT

*Over the past few years, emergence of Cloud Computing has notably made an evolution in the IT industry by putting forward an 'everything as a service' idea .Cloud Computing is of growing interest to companies throughout the world, but there are many barriers associated with its adoption which should be eliminated. This paper aims to investigate Cloud Computing and discusses the drivers and inhibitors of its adoption. Moreover, an attempt has been made to identify the key stakeholders of Cloud Computing and outline the current security challenges. A SWOT analysis which consists of strengths, weaknesses, opportunities and threats has also carried out in which Cloud Computing adoption for SMEs (Small and Medium-sized Enterprises) is evaluated. Finally, the paper concludes with some further research areas in the field of Cloud Computing.*

## 1. INTRODUCTION

Quite recently, considerable attention has been paid to Cloud Computing due to its inherent power so that transmute what happens in the heart of IT industry [1]. In Cloud Computing, new IT services come into view from the collaborative convergence of business and technology perspectives; moreover, enabling users to gain accessibility of explicit knowledge, services could effectively make a contribution to information and knowledge sharing [2].

In the light of cloud technology, practically it would be a good opportunity for SMEs; because they do not have a significant amount of resources and technical expertise to set up the appropriate infrastructure so as to compete with their larger competitors. It seems that the most dramatic advantage which Cloud Computing brings for SMEs, is to offer returns on investment that has never been possible before. Cloud Computing aims to reduce the amount of complexity, minimize costs, and enhance organizational agility [1].

Although there is a great potential for success, SMEs should be aware of the risks involved. Like other kinds of outsourcing, there is a strong need of precision and caution for enterprises to be informed of the inherent risks and have the capability of overcoming them [3].

Cloud Computing decrease the obstacles to conduct information process intensive activities; Indeed, people do not need to maintain their own technology infrastructure as they transfer the burden of system management and data protection to the cloud computer service provider [4,5].

Undoubtedly, Cloud Computing provides noticeable opportunities for enterprises; but more amount of time is needed to be taken for its development in the IT industry and in fact Cloud Computing is a newly introduced phenomenon yet.

The salient factor in Cloud Computing's adoption is the appropriate cost reduction structure which causes SMEs to be capable of launching and continuing their IT trends with the less need of high expenses [6].

The paper is organized as follows. Section 2 presents a variety of definitions toward cloud computing; section 3 introduces the main stakeholders of Cloud Computing; sections 4 and 5 are dedicated to investigate drivers and inhibitors of its adoption for SMEs and discuss about the affiliated advantages and drawbacks of it; section 6 explores some of security challenges; section 7 presents a SWOT analysis on Cloud Computing adoption for SMEs; section 8 concludes with a summary and finally section 9 outlines some issues for further and future study.

## 2. DEFINITION OF CLOUD COMPUTING

Cloud computing is defined as a novel model where it doesn't require user's ownership of necessary resources such as hardware and software, and instead the users can use them over the internet. A simple definition of Cloud Computing confirmed by many authors is: In Cloud Computing, the ownership, management upgrade and maintenance of resources are duties of third parties and end user's involvement is not needed. [7].

The technology of Clouds is hardware-based and hardware management is separated from the buyer .Cloud services deliver three kinds of capacity including computer, network and storage [8].

Vaquero in [9] reports that clouds are a large pool of readily, usable and accessible virtualized resources like hardware, development platforms and/or services. Such resources have the ability to be dynamically reconfigured to adjust to a changeable load (scale), allowing further for an optimized resource utilization. This pool of resources is typically extracted by a pay-per-use model in which guarantees are offered by the Infrastructure Provider through customized Service Level Agreements".

Buyya in [10] expresses an alternative definition as follows: "Cloud is a parallel and distributed computing system comprising a set of inter-connected and virtualized computers that are dynamically provisioned and presented as one or more unified computing resources based upon service-level agreements (SLA) developed via negotiation between the service provider and consumers".

According to National Institute of Standards and Technology (NIST) definition "the Cloud Computing is a model for providing easy access based on end user's requirement through Internet to set of configurable cloud resources (networks, servers, application, storage capacities) in such a way that the access can be provided expeditiously without intense need of resources management or direct involvement of service provider. NIST presents another scope of basic services that are provided by Cloud Computing, which involve software, platform, and infrastructure [11].

### 2.1. Cloud Computing Categorization

Cloud Computing is typically categorized on either its deployment or service models. Moreover, deployment models of clouds and cloud service models are listed in Tables 1 and 2, respectively [11].

Table1. Deployment Models of Clouds

| | |
|---|---|
| **Public cloud** | Its ownership is by a service provider and all the resources are accessible publicly. End users have the option of renting and adjusting resources based upon their own consumption pattern. |
| **Private cloud** | Is owned or rented by an organization for its private use. |
| **Community cloud** | It is similar to Private cloud but here resources are shared between members of a closed group who have the same needs. |
| **Hybrid cloud** | Combination of two or more cloud infrastructures (which can be public, private or community) /provide extra resources in cases of high demand. |

Table2. Cloud Service Models

| | |
|---|---|
| **SaaS** | This service is a software model depends on stored data in cloud environment and also a kind of technology providing possibility of access to software remotely. It can be a shared model to deliver most of business applications such as ERP, CRM, and HRM. [12] |
| **PaaS** | This service delivers a software layer in a package mode which can be used in generating higher levels services. The applications are developed and acquired by end users on top of the tools provided. Google Apps is an example of PaaS. |
| **IaaS** | This service is the use of fundamental computing resources, e.g. Storage, network, servers to provide services to end users. |

## 2.2. Features of Cloud Computing

Cloud Computing provides a compelling value proposition for organizations to outsource their Information and Communications Technology (ICT) infrastructures [13].

Miller[11], [14] proposes that Cloud Computing is user-centric and task-centric, and distributed computing can provide more effectiveness for sharing resources and collaborations in a group. The report of NIST further presents five essential characteristics of Cloud Computing, which are on-demand self-service, broad network access, resource pooling, rapid elasticity, and measured service. Other features of cloud computing are listed in Table 3.

Table3. Features of Cloud Computing

| Parameters | Cloud Computing |
|---|---|
| Access | Via web |
| virtualization | Essential |
| Switching cost | High, due to incompatibilities |
| Ease of use | Easy |
| Business model | Pricing (based on rent rate) |
| Application development | In the cloud |
| control | Centralized |
| openness | Low |
| Service level agreements | Essential |

## 3. KEY STAKEHOLDERS OF CLOUD COMPUTING

Despite the traditional approach that just has two major groups of stakeholders encompassing service providers and consumers, in Cloud Computing stakeholders are more detailed. These stakeholders summarized in Table 4, included two other groups which are regulators and enablers, in addition of the traditional ones.

In Cloud Computing, role of different traditional stakeholders will be altered. In the traditional computing, consumer is responsible for maintenance and upgrade processes while in Cloud Computing approach those processes are parts of provider's duties.

Table 4. New Stakeholders of Cloud Computing

| | |
|---|---|
| providers | Vendors who Perform the maintenance and upgrade of the system and are responsible for protecting consumer's data. Providers include well established companies such as Google, Microsoft, IBM, Oracle, Amazon, Sun, 10Gen, Salesforce, and Dataline to newer companies like Netsuite Corporation, Appistry, AppRiver, Boomi, Carbonite, and Enomaly. |
| consumers | Subscribers who purchase and make use of system |
| regulators | International entities that permeate across the other stakeholders |
| Enablers | Organizations that are responsible for selling services, facilitating the delivery, adoption and utilization of cloud computing |

## 4. KEY DRIVERS OF CLOUD COMPUTING ADOPTION IN SMEs

Making use of a cloud service will lead to lower capital investments and required costs, yet services are provided in real-time; besides, as mentioned earlier, vendors become responsible for all the maintenance tasks including updating and upgrading. According to Rayport and Heyward [15], major drivers of cloud services are as below: anywhere / anytime accessibility to cloud based software, cloud enabled storage as a ubiquitous service, specialization and customization applications, collaboration among users and cost advantage predicted on cloud efficiencies, warehouse-size data centers, energy efficiency and everything as a service. Moreover, other drivers of cloud services are detailed as follows:

- **Cost Reduction Structure**: Cloud computing approach remarkably decreases the cost of entry for SMEs which trying to benefit from compute-intensive business analytics that were hither to available only to the largest of companies. As SMEs lacked the enough necessary resources and also cannot afford the intensive expenses, Cloud Computing can provide numerous opportunities for them. Moreover, it can enable developing countries through a good cost reduction structure to compete more in business world and to propose their new IT solutions globally.
- **Quick Accessibility**: Cloud Computing can present an almost urgent access to hardware resources that not only doesn't require much upfront capital investments for users, but also results in a faster time to market in many businesses. The users are thoroughly separated from each other.
- **Innovation Incentive Structure**: Cloud Computing can play a significant role in encouraging innovation and reducing barriers related to innovation in IT field through social networks such as Facebook and twitter.

- **On Demand Structure**: Cloud computing makes it easier for SMEs to scale their services according to client demand. As they notify to need renting new resources, the required resources are delivered quickly through the internet [16].

## 5. KEY INHIBITORS OF CLOUD COMPUTING ADOPTION IN SMEs

Albeit there are numerous benefits to adopting cloud computing, there are also some significant barriers associated with its adoption that would be addressed. Obviously, security is one of the most crucial concerns in the adoption of Cloud Computing so that lack of enough security can result in hindering the adoption process for SMEs; However, there are some solutions offering by cloud providers which can maximize the security level convincingly such as User-centric IdM and Federated IdM solutions . After security , outage (temporary loss of service), interoperability (portability or ability to change supplier ), and the reliability are the most significant ones [17] , [18].

The inhibitors of cloud computing's adoption are detailed as follows:

- **Reliability:** Enterprise applications are presently so significant that they must be reliable and available to support operations. In the case of failure, recovery plans must begin with minimum disruption. Further costs may be pertaining to the essential levels of reliability; nevertheless, the business can do merely so much to decrease risks and the failure cost. Developing a track record of reliability will be a necessity for the extensive adoption.
- **Connectivity and open access**: The full realization of Cloud Computing is dependent on the availability of high-speed access to all. In fact, open access to computing resources should be similar to water and electricity power accessibility.
- **Security and Privacy:** As Cloud Computing approach presents a novel delivering model of IT solutions, it seems that there is a relatively high rate of insecurity associated with its entrance towards the business world. It is noticeable that there is lack of enough certainty and security accompanied with the advent of every new technology or innovation and it is inevitable. The capability of Cloud Computing to sufficiently address privacy regulations has been called into question. These days, organizations come across to multitudinous different requisites making effort to safeguard the privacy of individuals' information.
- **Interoperability:** In adoption process of Cloud Computing by SMEs , it is vital to exist an appropriate level of interoperability between public and private clouds.  A large number of companies have made remarkably progress toward standardizing their processes, data, and systems by means of implementation of ERPs. Standardization requires an extensible infrastructure resulting to a fully integrated connection among instances. SaaS applications delivered via the cloud procure a low-capital, fast-deployment option. Depending upon the application, it is significant to integrate with traditional applications that may be resident in a separate cloud or on traditional technology. Standard can be as an enabler or an impediment for interoperability that both of them yield a good maintenance of information and process's integration. [19]
- **Economic Value:** It sounds intuitive that by sharing resources to smooth out peaks, paying solely for what is used, and cutting upfront capital investment in employing IT solutions, the economic value will be there.  There will be a necessity to accurately balance all costs and benefits relevant to Cloud Computing—in both the short and long runs. Hidden costs could encompass support, disaster recovery, application modification, and data loss insurance. Since usage expands and interoperability requirements for the business process become more onerous, a novel approach is required.
- **Political Issues Due to Global Boundaries:** In the world of Cloud Computing, there is variability in terms of where the physical data resides, where processing takes place, as well

17

International Journal of Information Technology Convergence and Services (IJITCS) Vol.4, No.2, April 2014

as from where the data is accessed. Given this variability, various privacy rules and regulations may utilize. Obviously, due to the variable nature of rules and regulations, politics and direct involvement of government become more salient in the adoption process of Cloud Computing. For Cloud Computing to constantly evolve into a borderless and global tool, it necessitates to be separated from politics. Presently, some significant global technological and political powers are making laws that can have an inverse impact on development of the cloud solutions globally [18]. Providers have been unable to assure the location of a company's information on specific set of servers in a specific location. Nevertheless, cloud computing service providers are swiftly employing measures to address this issue. Briefly toward politic issues, Cloud Computing is highly dependent on global politics to survive. Providing open access connectivity and enough bandwidth and also allocating enough capital for ICT field are all under the control of government and influenced by the politics.

## 6. CLOUD SECURITY CHALLENGES

- **Multi tenancy issue**: It means the ability of using the same software and interfaces to configure resources and isolate customer- specific traffic and data. This issue seems like a sensor which is sensitive to any unauthorized access from other users which are running processes on the same physical servers.
- **Authentication of acquired information**: When cloud user rents cloud software, all the data are in the cloud provider's servers under the control of him and user doesn't have any access to it. Hence, there is the possibility of making any change in the user's data without his permission. Therefore, the authentication of the data in this case is very crucial, and therefore needs to be guaranteed [19].
- **Resource location** : End users use the services offered by the cloud providers without by means of being informed about the accurate location of resources .This could cause a potential problem which is sometimes over the control-domain of cloud providers [20].
- **System monitoring and logs** : As more business critical applications are transformed to the cloud environment , customers may call for more monitoring and log data from providers for their personnel.
- **Cloud standards** : standards are essential across different standard developing organizations to gain interoperability among clouds and to enhance their solidity and security [21].

## 7. SWOT ANALYSIS FOR ADOPTING CLOUD COMPUTING SERVICES IN SMES

The acronym SWOT stands for Strength, Weaknesses, Opportunities, and Threats. SWOT analysis is an efficient tool used to identify environmental conditions and intra organizational capabilities involved in every project and has been used extensively in various decision making processes. In this analysis, firstly the goal of project and secondly its internal and external determinant factors are identified. This method can make a contribution in investigating and evaluating issues from all main aspects in which every issue is analyzed comprehensively based on the mentioned factors [13].

In order to assess the adoption of Cloud Computing for SMEs in a more comprehensive manner, a SWOT analysis is conducted. The results of the analysis are summarized in Figure 1.

18

| Internal | |
|---|---|
| **Strengths** | **Weakness** |
| 1. Cost effective<br>2. Flexible and innovative<br>3. Simplified cost and consumption model<br>4. Faster provisioning of systems and application<br>5. Secured infrastructure<br>6. Compliant facilities<br>7. Resilient in disaster recovery<br>8. Maintenance Cost Reduction<br>9. Convenient level of accessibility<br>10. Butter control of the resources<br>11. Independence of time and location<br>12. Energy saving<br>13. Environmental protection<br>14. Friendly utilization<br>15. expandability | 1. Post training required<br>2. Development of applications<br>3. Increased dependency<br>4. High- speed Internet connection requirement<br>5. Difficulty of integration with local software<br>6. Data transfer bottlenecks<br>7. Lack of physical control of data<br>8. Lack of commitment to the highquality of service and availability and availability guarantees<br>9. Inability of providers to guarantee The location of the company's information |
| **Opportunities** | **Threats** |
| 1. Pay for use licenses<br>2. Good chance for SMEs because of making progress without upfront investments<br>3. Invent scalable store<br>4. Marketplace enhancement in terms of functionality, innovation & price<br>5. Adaptive to future needs<br>6. Standardized process<br>7. Quick solution of the problem<br>8. High-tech work environment<br>9. Offering modern information solutions according the last technology | 1. Security concerns (data security)<br>2. lack of specific standard regulation (local, national & international)<br>3. Difficulty from migration from one to another platform<br>4. Hidden cost (backup, problem solving and recovery)<br>5. compatibility reduction<br>6. Possibility of backlash from entrenched incumbents |
| External | |

(Positive on left side, Negative on right side)

Figure 1. SWOT Analysis for Cloud Computing

## 8. CONCLUSION

In this paper we tried to identify the key stakeholders of Cloud Computing and investigate the different issues associated with Cloud Computing as well as advantages and disadvantages of its adoption. To achieve this, a SWOT analysis was accomplished in which strengths, weaknesses, opportunities as well as threats of cloud computing adoption was appraised.

The results of the analysis indicated that SMEs can make progress dramatically in their business trends through cost reduction structure and faster software upgrade. Lower investment on infrastructure and hardware, easier scale up of applications and more efficient use of computing resources are other benefits of cloud computing solutions for SMEs. Also, a Monthly - based fees structure is a significant factor for SMEs to adopt cloud approach. Nevertheless, Cloud Computing services are not free of risks. There is a real risk of the lack of information and system security if proper actions are not taken to safeguard information and system security. Also, there is usually some amount of ambiguity for cloud users about the accurate place of data and level of its privacy.

Therefore, deploying strategies toward providing more security can directly influence on raising the adoption rate.

## 9. FURTHER RESEARCH AND FUTURE WORK

Admittedly, in order to extend our knowledge in this field, further research on investigating adoption of cloud computing in SMEs is essential. In our further research, we intend to concentrate on PEST analysis so as to investigate effective factors in adoption of Cloud solutions. In addition, regarding the fact that security issues are one of the most crucial concerns which inhibit cloud computing adoption, further research can be dedicated to concentrate deeply on security challenges and explore some solutions to tackle the barriers.

## REFERENCES

[1] Kochut, A., 2011 "Evolution of the IBM cloud: enabling an enterprise cloud services Ecosystem", IBM J. Res .Dev ., 55, 6.
[2] Amato, F.; Mazzeo, A.; Moscato, V.; Picariello, A., 2014 "Exploiting Cloud Technologies and Context information for recommending touristic paths", In Intelligent Distributed Computing VII, Springer, pp. 281–287.
[3] Vouk, M.A., 2008 "cloud computing –issues, research and implementations", Journal of Computing and Information Technology.
[4] Jaeger, P.T.; Lin .J. Grimes, J.M., 2008 "Cloud computing and information policy: computing In a policy cloud?", journal of information technology & politics.
[5] Christensen, C.M.; Anthony, S.D.; Roth, E. A., 2004 "Seeing what's next: using theories of Innovation to predict industry change", Boston,MA :Harvard Business School Press.
[6] Staten, J., 2009 "Hollow out the MOOSE: Reducing cost With Strategic Right sourcing", Forrester Research, Inc.
[7] Iyer, B.; Henderson, J. C., 2010 "Preparing for the future— understanding the seven capabilities Of Cloud Computing", MIS Quarterly Executive, pp. 117–131.
[8] McKinsey & Co., , 2009 "Clearing the Air on Cloud Computing", Technical Report.
[9] Vaquero, LM; Rodero-Merino, L.; Caceres J.; Lindner, M., 2009 "A break in the clouds: Towards a cloud definition", SIGCOMM Computer.
[10] Buyya, R.; Yeo, CS.; Venugopal, S.; Broberg, J.; Brandic, I., 2009 "Cloud computing and Emerging IT platforms: Vision, hype, and reality for delivering computing as the 5th utility", Future Generation Computer Systems, 25, pp. 599-616.
[11] National institute of Standards and Technology, 2011. "The NIST definition of cloud Computing" .
[12] Cusumano, M., , 2010 "Cloud computing and SaaS as new computing platforms", Communications of the ACM.
[13] Haynie, M., 2009 "Enterprise cloud services: Deriving business value from cloud computing", Micro Focus, Tech. Rep.,.
[14] Miller, M., , 2008 "Cloud Computing: Web-Based Applications That Change the Way You Work and Collaborate Online", Indianapolis, Indiana: QUE.
[15] Rayport, J.F; Heyward, A., 2011 "Envisioning the cloud: The Next Computing Paradigm And Its Implication For Technology Policy".
[16] Dubey, A.; Wagle, 2007 "Delivering software as a service", The Mckinsey Quaterly.
[17] Voorsluys, W.; Broberg J.; Buyya, R., 2011 "Cloud Computing Principles and Paradigm", John Wiley And Sons.
[18] Zissis, D.; Lekkas, D., 2012. "Addressing cloud computing security issues", Future Generation Computer Systems, 28(3), pp.583–592.
[19] ]AlZain, M. A.; Pardede, E.; Soh, B.; Thom, J. A, , 2011 "Cloud Computing security: From single to multi-clouds", In Proceed-ings of the HICSS, pp. 5490–5499.
[20] Chunming, R.; Nguyen Son, T., 2011 "Cloud trends and security challenges", In: Proceedings of the 3rd international workshop on security and computer networks (IWSCN 2011).
[21] Karin, B.; Gilje, JM.; Håkon, MP., 2011 "Undheim Astrid. Security SLAs for federated cloud services", In: Proceedings of the 6th international conference on availability, reliability and Security.